\documentclass[preprint]{aastex631}
\usepackage{xspace}
\usepackage{graphicx}
\usepackage{natbib}
\usepackage{url}
\usepackage{bm}
\usepackage{amsmath}
\usepackage[maxfloats=256]{morefloats}

\newcommand{\degree}{\ensuremath{\,^\circ}}
\newcommand{\degper}{\rlap.{^{\circ}}}
\newcommand{\arcmper}{\rlap.{^{\prime}}}

\newcommand{\mk}{\ensuremath{\,{\rm mK}}}

\newcommand{\kpc}{\ensuremath{\,{\rm kpc}}}

\newcommand{\kms}{\ensuremath{\,{\rm km\, s^{-1}}}}

\newcommand{\hi}{H\,{\sc i}}

\newcommand{\hii}{H\,{\sc ii}}

\newcommand\urltilda{\kern -.15em\lower .7ex\hbox{\~{}}\kern .04em}
\newcommand{\gsim}{\ensuremath{\gtrsim}}

\begin{document}

\title{The Flocculent Structure of the Inner Milky Way Disk}

\author[0000-0002-2465-7803]{Dana S. Balser}
\affiliation{National Radio Astronomy Observatory, 520 Edgemont Rd.,
  Charlottesville, VA 22903, USA}
\email{dbalser@nrao.edu}

\author{W. B. Burton}
\affiliation{Sterrewacht Leiden, The Netherlands}
\email{wbutlerburton@gmail.com}

\begin{abstract}

  Observations of \hi\ published in 1957 by Westerhout and Schmidt
  were presented as showing a global face-on view of spiral structure
  in the Milky Way.  Since then many studies have attempted to improve
  on the early map, perhaps presupposing our Galaxy to be
  characterized by a Grand-design pattern of prominent spiral arms.
  We consider here two approaches to explore the nature of spiral
  structure of the inner Milky Way disk using the HI4PI survey.  The
  first is to search for shallow shoulders in the high-velocity wings
  of \hi\ data along the Galactic equatorial disk of the inner Milky
  Way that would be expected if the lines of sight swept across
  interarm regions of low \hi\ density.  The second is to look for
  broad dips in the integrated \hi\ brightness temperature over the
  high-velocity wings, pertaining to gas near the subcentral region,
  that would be expected for the interarm region of a Grand-design.
  We find neither shallow shoulders nor broad dips in either the
  Northern quadrant I or the Southern quadrant IV indicating that the
  Milky Way seen interior to the Solar orbit is not characterized by a
  majestic spiral-structure Grand-design; this conclusion is a robust
  one, in that it does not depend on measures of distance.  Taken
  together with decades of work on the bits and pieces of the quite
  disorganized shambles of the inner Galaxy, we suggest that the Milky
  Way belongs to the category of Flocculent spirals.

\end{abstract}


\section{Background}\label{sec:background}

Many decades of work have focused on improving in detail, using modern
data of various sorts, on the \hi\ map of the Milky Way visible from
the North, made with the Kootwijk 7-m telescope by
\citet{westerhout57} and \citet{schmidt57}, soon combined with data
from the South by \citet{kerr57}.  Yet much of that following work
seems to have been aimed at mapping the spiral structure of the Milky
Way that was presupposed\footnote{The dictionary definition of
  ``Presupposed'' seems relevant to what motivated many efforts to
  determine the global morphology of the Milky Way: ``...tacitly
  assume at the beginning of a line of argument or course of action
  that something is the case.''}  to be characterized by a
Grand-design pattern, with, for example, measurable tilt angles of
logarithmic spirals of a measurable number wrapping over large angles
of Galactocentric azimuth.

A percipient remark was made by \citet{shane66}.  They wrote that, if
the inner Galaxy morphology were characterized by clearly defined
spiral arms, in other words by clearly defined interarm regions, then
the spectra observed transiting an interarm region near the subcentral
points\footnote{The subcentral points are the locus of points where
  the lines of sight are closest to the Galactic center, and where the
  radial velocity would be greatest in the case of circular
  velocity.}, where the spiral loop of the interarm region would be
seen tangentially and at its highest velocity, would show a weak,
shallow shoulder.  Quoting them: ``{\it The hydrogen density may, in
  fact, be so low in the tangent region that only a low-lying
  extension to the line profile is visible.  No clear case of this has
  been observed...}[emphasis added].''  In other words, no clear case
of an interarm region has been observed.  This quote is from page 272
of their lengthy article which focused on an effort to define the
Galactic rotation curve, rather than on determining the global
morphology.  \citet{shane66} found no clear shallow shoulders in their
spectra but suggsted that a shallow shoulder might explain the
spectrum at $\ell = 68$\degree\ in their Figure~6.  Interarm gas near
the subcentral region with velocities extending just below the
terminal velocity would be blended with brighter emission from the arm
region.  In such cases the blended profile would be broadened.

But \citet{burton71} did follow up on the percipient remark, by
measuring the sharpness of the high-velocity cutoff on \hi\ profiles
measured along a portion of the quadrant I Galactic equator, from
$\ell = 22$\degree\ to $\ell = 56$\degree, the range then available in
the data from the Dwingeloo 25-m telescope. The resulting velocity
dispersion, $\sigma$, of the Gaussian fits to the edges is shown in
Figure~3 of the 1971 paper.  That plot gave quantitative support to
the remark of \citet{shane66}: no shallow shoulders with high $\sigma$
in the edges of the profiles that would be consistent with the lines
of sight sweeping across an interarm region are observed.  The
conclusion could have been more directly stated, that if the inner
Galaxy showed no dominant {\it interarm} regions, then it shows no
spiral {\it arm} regions either.  Yet just as \citet{shane66} did not
take the consequences of their remark, neither did \citet{burton71}
follow up directly on the results of his Figure~3.

\section{Data and Analysis}\label{sec:data}

We use here the \hi\ $4\pi$ survey (HI4PI)\footnote{See
  https://cdsarc.cds.unistra.fr/viz-bin/cat/J/A+A/594/A116}, an
all-sky database of Galactic \hi\ with an rms sensitivity of
43\mk\ and an angular full-width at half-maximum (FWHM) resolution of
$16\arcmper2$ \citep{hi4pi}, to investigate the conclusions of
\citet{shane66} and \citet{burton71} regarding the lack of strong
evidence for interarm regions of spiral structure with modern
\hi\ data.  This survey consists of observations made with the
Effelsberg 100-m and Parkes 64-m telescopes in the Northern and
Southern hemispheres of the Galactic disk, respectively.

We generate abridged spectra using the Cube Analysis and Rendering
Tool for Astronomy (CARTA 4.1.0)\footnote{See https://cartavis.org/.}.
Circular regions are created on the sky from Galactic longitude of
$\ell = 20^{\circ}-80$\degree\ in the North and $\ell =
280^{\circ}-340$\degree\ in the South, both in increments of 1\degree.
All regions are centered in the plane of the Milky Way ($b =
0$\degree), with a diameter of 1\degree.  The 122 spectra are averages
over these regions.

We fit a Gaussian distribution to the high-velocity shoulder of the
spectral profiles near the subcentral point using the SciPy
optimization program {\tt curve\_fit}.  The velocity range used to fit
the data was specified by eye, together with the estimated amplitude,
center, and width of the Gaussian profile.  The width is defined as
the standard deviation, $\sigma$.

\begin{figure}
  \centering
  \includegraphics[angle=0,scale=0.55]{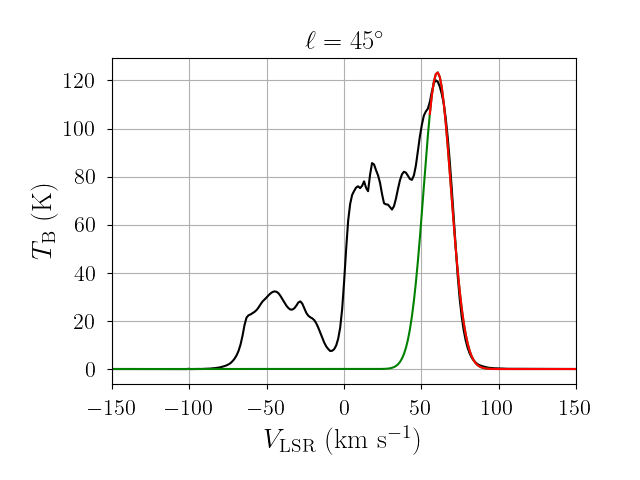}
  \includegraphics[angle=0,scale=0.55]{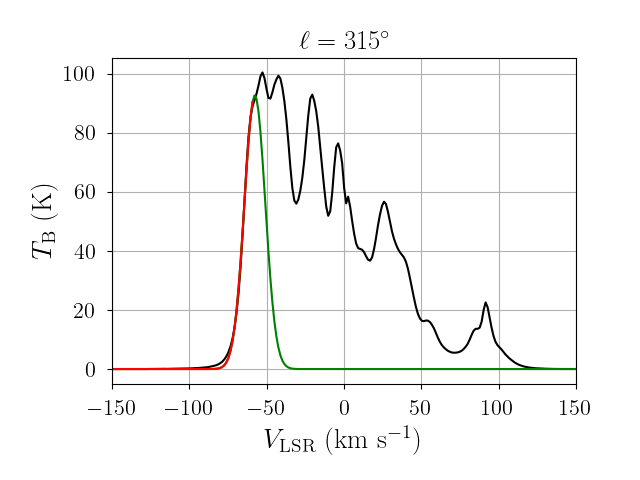}
  \caption{Representative HI4PI \hi\ spectra for $\ell =
    45$\degree\ (left) and $\ell = 315$\degree\ (right).  The green
    curve is the Gaussian fit of the high-velocity shoulder and the
    red curve shows the portion of the \hi\ data that was used to
    constrain the fit. The shape of the extreme-velocity shoulder is
    measured by the dispersion, $\sigma$, of the Gaussian fit.}
  \label{fig:spectra}
\end{figure}

Figure~\ref{fig:spectra} shows representative \hi\ spectra, for $\ell
= 45$\degree\ and $\ell = 315$\degree.  The strategy is to fit a
Gaussian to the extreme velocity peak, or near that peak, such that
the resulting Gaussian fits the shoulder well.

\section{Results}\label{sec:results}

Following \citet{burton71}, we look for properties of the morphology
of the Milky Way by determining the sharpness of the high-velocity
cutoff measured by the Gaussian fits and by exploring fluctuations in
the total integrated \hi.  New here is that we also measure the total
integrated \hi\ over just the high-velocity wing as defined by the
Gaussian fit.

\begin{figure}
  \centering
  \includegraphics[angle=0,scale=0.55]{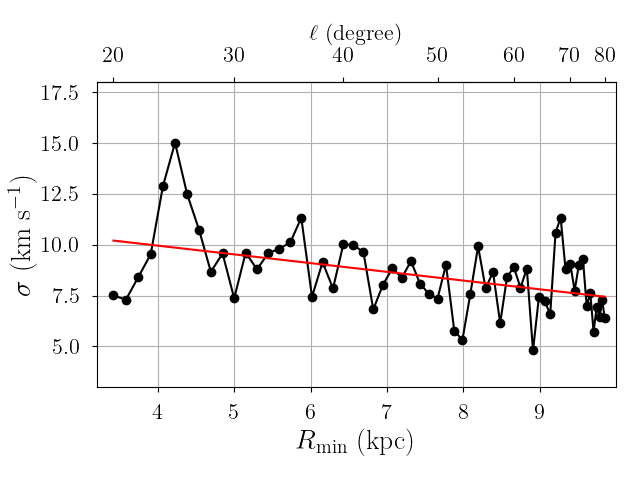}
  \includegraphics[angle=0,scale=0.55]{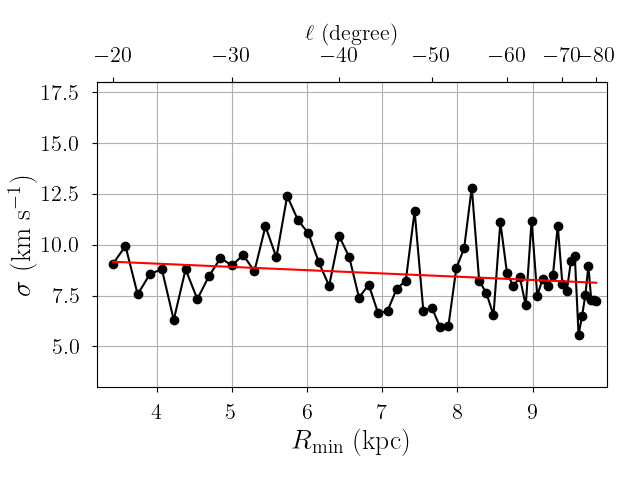}
  \caption{Apparent velocity dispersions as a function of the
    Galactocentric radius of the subcentral point for the Northern
    data (left) and Southern data (right).  The red curve is a linear
    fit to the data, yielding $\sigma = -0.43\,R_{\min} +
    11.67$\kms\ for the Northern data and $\sigma = -0.16\,R_{\min} +
    9.71$\kms\ for the Southern data.  The units of $R_{\rm min} =
    R_{\rm o}\,{\rm sin}(\ell)$ are in kpc and we assume $R_{\rm o} =
    10$\kpc\ (see text). The measurements show no high $\sigma$
    excursions that would be expected from relatively empty interarm
    regions. }
  \label{fig:sigma}
\end{figure}

Figure~\ref{fig:sigma} shows the standard deviation of the Gaussian
fits discussed above plotted as a function of the Galactocentric
radius of the subcentral point, $R_{\rm min}$ \citep[see Figure~3
  of][]{burton71}.  We chose a Galactocentric distance of $R_{\rm o} =
10$\kpc\ to facilitate comparison with earlier plots, but this choice
does not affect the conclusions.  The apparent dispersion is somewhat
higher than that determined by \citet{burton71} using less detailed
data, and spans a slightly larger range, but otherwise the trend is
similar.  The red curve is a linear fit to the data and produces a
similar slope as estimated by \citet{burton71}.

\begin{figure}
  \centering
  \includegraphics[angle=0,scale=0.55]{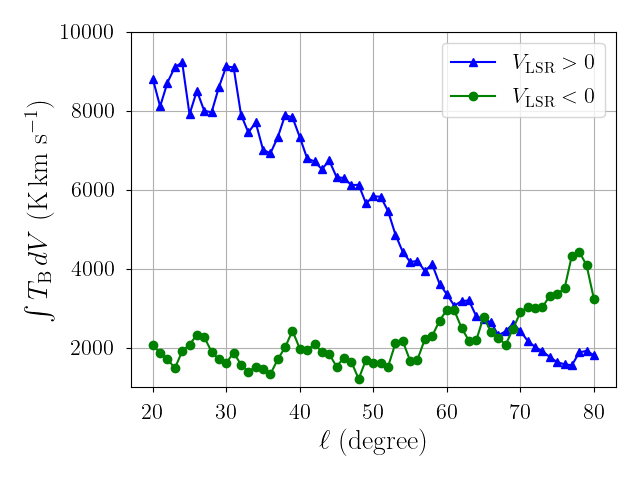}
  \includegraphics[angle=0,scale=0.55]{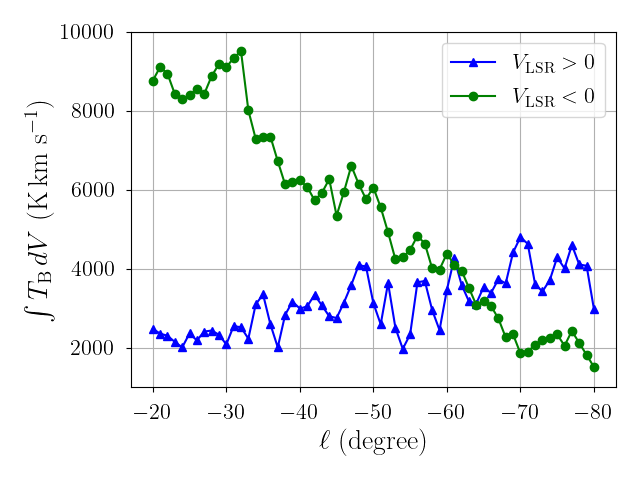}
  \includegraphics[angle=0,scale=0.55]{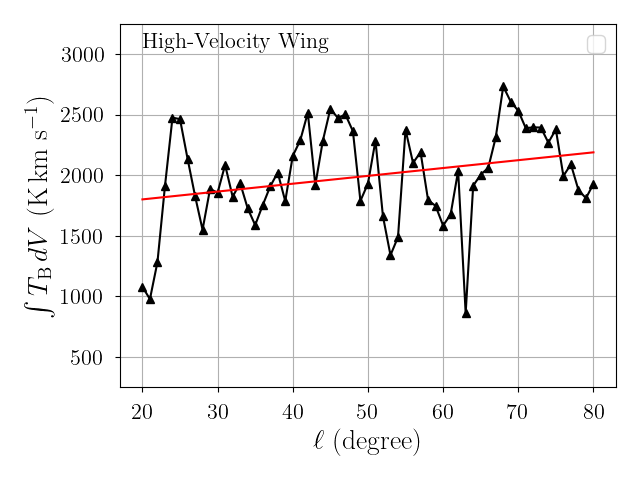}
  \includegraphics[angle=0,scale=0.55]{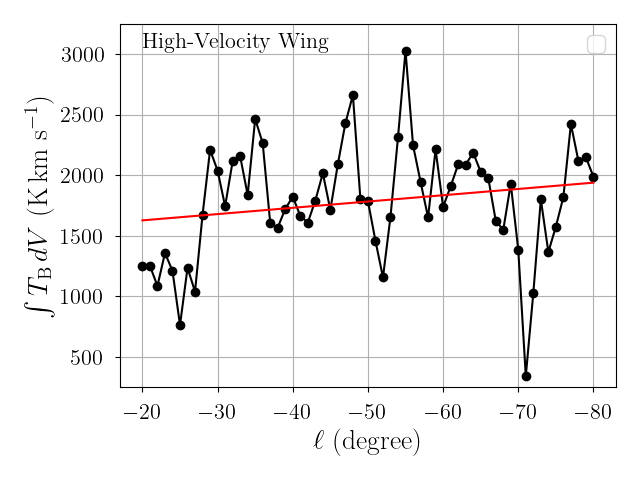}
  \caption{Integrated brightness temperatures as a function of
    Galactic longitude for the Northern data (left) and Southern data
    (right). {\it Upper:} The brightness temperature integrated over
    all LSR velocities: the positive and negative LSR velocities are
    shown independently, representing the inner and outer parts of the
    Galactic disk. The hills and valleys that would be expected if a
    dominant spiral pattern pertained in the inner Galaxy are
    absent. The North-South differences are no doubt due to the
    slightly lopsided nature of the Milky Way, a property not
    uncommon. {\it Lower:} The brightness temperature integrated only
    over the high-velocity wings, pertaining to gas near the
    subcentral region.  The red curve is a linear fit to the data,
    yielding $\int{T_{B}\,dV} = 6.48\,\ell + 1672\,{\rm K}\,{\rm
      km}\,{\rm s}^{-1}$ for the Northern data and $\int{T_{B}\,dV} =
    -5.19\,\ell + 1524\,{\rm K}\,{\rm km}\,{\rm s}^{-1}$ for the
    Southern data.  There are no broad dips that would be expected for
    the interarm region of a Grand-design. We comment in the text on
    the apparently ``noisy'' nature of the lower plots.}
  \label{fig:bint}
\end{figure}

Figure~\ref{fig:bint} shows the integrated brightness temperatures as
a function of Galactic longitude \citep[see Figure~12 of][]{burton71}.
Here, we sum the product of $T_{\rm B} * \Delta{V_{\rm LSR}}$ from
$-200$\kms\ to 0\kms\ for negative velocities, and from 0\kms\ to
$+200$\kms\ for positive velocities.  If the interarm region is
tangent to the line-of-sight then the total integrated \hi\ over all
velocities would be lower: a valley.  The detailed fluctuations are
different than those shown in \citet{burton71} based on more primitive
data, but the general trends are similar.  The hills and valleys that
would plausibly be expected as the lines of sight swept the inner
Galaxy, if the Galactic morphology was characterized by a clear
spiral-arm signature, are absent.

We also integrate the brightness temperature over only the
high-velocity wing, corresponding to gas near the subcentral region.
We do this by integrating from 10\kms\ removed from the terminal
velocity, defined by the center of our Gaussian fit, to 200\kms\ for
the Northern data and --200\kms\ for the Southern data.  The higher
spectral resolution and sensitivity of the HI4PI survey is more likely
directly to detect the shallow shoulder from an interarm region, thus
producing a dip in the integrated emission near the subcentral region.
The results in the bottom panels of Figure~\ref{fig:bint} show
fluctuations in the integrated brightness temperature but not broad
dips that would be expected of thick arm and interarm regions for a
Grand-design.  The ``noise'' in the plot is expected given the
activity near the locus of subcentral points, since several kpc of
length are sampled in the velocity band of somewhat more than 10\kms.
For example, the narrow dips at $\ell = 63$\degree\ and $\ell =
-71$\degree\ can be identified with the remnant of a large-scale
explosive event first explored by \citet{katgert69} and the Carina
Flare supershell studied by \citet{dawson08}, respectively.  The
slight increase in integrated brightness as the line of sight sweeps
to larger longitudes is likely the result of more velocity crowding at
the higher longitudes which provides a longer path length near the
subcentral locus \citep[see Figure 3 of][]{burton71}.

\section{Discussion}\label{discussion}

Spiral galaxies take a wide range of forms.  Some are classified as
Grand-design systems if clearly-defined spiral arms wind outward from
the inner reaches over many degrees of Galactocentric azimuth; many
others are classified as Flocculent galaxies if the global morphology
is dominated by a muddle of apparently disconnected individual
features, or by low contrast or irregular arms
\citep[e.g.,][]{sarkar23}.  The morphology provides insight into
galaxy physics; for example, Flocculent spirals can be understood by
random local instabilities \citep{elemgreen11}.  There is clear
evidence of spiral structure in the early \hi\ Galactic disk maps in
the outer Galaxy where the situation is less confusing: no velocity
crowding near the subcentral point \citep{westerhout57}.  Westerhout
described properties of the outer Galaxy disk that remain largely
unchallenged by more modern data \citep[see][]{koo17}. Thus the {\it
  outer} Galaxy disk is characterized by quite clearly defined spiral
features that extend to large distances from the Galactic center; the
outer Galactic disk was also shown to be flared and warped.  But to
fully characterize the Milky Way requires tracing morphology over the
inner Galaxy as well.

Over the past decades there have been many studies, using various
tracers, attempting to map the anticipated spiral structure of the
full Galactic disk.  A full bibliography of relevant work would have
hundreds of entries: here we only mention some representative
works. The review by \citet{carraro15} accurately states that ``...a
coherent and clear picture is still missing.''  We emphasize that most
studies presuppose a Grand-design spiral or at least well-defined
spiral arms \citep[e.g.,][]{vallee22}.

\hii\ regions are a canonical tracer of spiral structure.  This was
recognized by Morgan who attempted to trace out spiral arms in the
neighborhood of the Sun \citep{morgan52}.  Morgan's complete earlier
work was never published, however, and this work became overshadowed
by the \hi\ maps that were not affected by extinction from dust.
After the discovery of radio recombination lines (RRLs) in 1965
\citep{hoglund65}, large RRL surveys detected many \hii\ regions
throughout the Galactic disk; for a review see \citet{wenger19}.  The
current census is that there are $\gsim 7,000$ \hii\ regions in the
Milky Way with $\gsim 2,500$ bonafide \hii\ regions that have RRL
detections \citep{armentrout21}.  The known \hii\ region sample
consists of nebula that are estimated to be ionized by at least one
O-type star and therefore should be relatively bright.  But attempts
to use these modern \hii\ region samples to produce a face-on map of
the spiral structure of the Milky Way \citep[e.g.,][]{hou14} are
hampered by streaming motions and the distant ambiguity in the inner
Galactic disk; and therefore, their results are not robust.

A small subset of the \hii\ region sample have parallax-determined
distances based on molecular maser emission associated with
\hii\ regions \citep[e.g.,][]{bian24}.  But in these studies only some
dozens of sources are used to define the presupposed spiral arms: they
do not provide clear evidence for any contrast expected from arm and
interarm regions.  Other tracers have also been used in attempts to
map structure in the inner Galaxy.  Following \hi\ studies, much
effort was directed to studies of the carbon monoxide distribution, as
CO serves as a surrogate for the seeds of star-formation: the survey
material compiled by \citet{dame01} has served as background material
for quite a few investigations of the global morphology.
 
Related studies have also included the information revealed by stellar
source counts from mid-infrared emission \citep{benjamin05}; a three
dimensional dust map using Gaia and LAMOST data \citep{wang25}; and
large-scale filaments ("bones") traced by molecular gas and dust
\citep{zucker18}.  \citet{peek22} demonstrate the difficulties in
interpreting large-scale structures in the inner Galaxy with similar
data sets.

Over the past decades many studies of Galactic spiral structure have
focused on \hii\ regions or molecular gas, thought to be more directly
associated with star formation.  But \hi\ comes before \hii\ and CO
and could be considered the primordial spiral arm tracer.  In many
spiral galaxies, including our own, \hi\ arms extend to great
galactocentric distances, well beyond the regions of active star
formation \citep[e.g.,][]{yidiz15}.  There is clear evidence of
prominent spiral arms in \hi\ in the outer Galaxy, we should therefore
be able to detect similar structures in the inner Galaxy.

We here avoid the problem of determining distances by focusing on the
properties of the \hi\ spectra near the locus of terminal velocities.
The measures of the shape of the \hi\ spectra at the cutoff velocities
of the subcentral region, shown in Figure~\ref{fig:sigma} separately
for the Northern and Southern portions of the Galactic equator, show
consistency, with no evidence for the shallow shoulders, i.e. higher
values of $\sigma$, that would be expected as the lines of sight swept
across interarm regions of low \hi\ density.  Applying in both
hemispheres, the conclusion is a general one; it is robust, not
depending on measurements of distances.

We note that the trend shown in Figure~\ref{fig:sigma} for the
Northern data affirms that shown in Figure~3 of \citet{burton71}, even
though the older figure was based on data much less complete and much
less sensitive. The differences between the two plots are not
significant to our results, but are perhaps worthy of comment.  The
older figure was based on a reduction by \citet{burton70} of entire
spectra into Gaussian components: thus neighboring Gaussians will each
contain a minor contribution from a tail of the neighbor; but in the
new figure, shown in Figure~\ref{fig:sigma} here, the Gaussian fit to
the high-velocity was done with a fit of a single Gaussian to the
high-velocity edge of the spectrum, without any contamination from
neighboring fits: thus the new fit is understandably somewhat sharper.
The newer data are also, of course, made with a much higher
velocity and angular resolution than the data from the Dwingeoo
25-m telescope. The Dwingeloo data used by Burton were mostly from
1963, with an effective angular resolution of $0 \degper 6$ and
velocity resolution of 1\kms; the full spectra in that work over the
full velocity range required many repeats stepping in velocity,
because the receiver measured only 8 [{\it sic}] channels at a time.

We also comment on a trend in both the new and old $\sigma - R_{\rm
  min}$ plots: the $\sigma$ values at the cutoff velocities tend to
decrease slightly with increasing $R_{\rm min}$.  This is seen in both
Galactic hemispheres as plotted in Figure~\ref{fig:sigma}.  Although
understanding this trend is not central to the conclusions now being
drawn, it is perhaps interesting to offer speculations as to why the
cutoff-velocity edges appear less sharp at larger $R_{\rm min}$ than
at smaller values.

The first speculation is a simple matter of geometry, and the
resulting velocity crowding.  Figure~11 of \citet{burton71} isolates
the effects of velocity crowding in the instructive, albeit na\"{i}ve,
case of a completely uniform distribution of \hi.  The structures in
this map near the terminal velocity become more intense in the
longitude range 70\degree--80\degree\ than in the range
20\degree--30\degree.  (A similar situation would pertain in the
South.)  Thus the contour lines are more tightly spaced at the higher
longitude range than at the lower one; the value of the Gaussian
$\sigma$ would consequently decrease with increasing longitude.

A second speculation, which we mention but with no further comment, is
more physical than the previous geometric one.  Could the Galaxy be
``hotter'' at the lower $R_{\rm min}$, leading to broader spectral
lines?  We note that star formation is more robust at lower $R_{\rm
  min}$ than it is closer to the Sun.

We also note that the slope of the $\sigma - R_{\rm min}$ plot for the
Northern data is somewhat steeper than that for the Southern.  In this
regard, we offer a third speculation, concerning the possible
influence of the bar on the measured shapes of the spectra
cutoffs. The Galactic bar is tilted with respect to the Sun-Center
line, under an angle of Galactocentric azimuth of some 30 or 40
degrees, such that the end of the bar is closer to the Sun in the
Northern hemisphere than in the Southern.  It would not be surprising
if the \hi\ lying near the bar would be stirred or heated to broader
widths.

Another way to look for evidence of a dominant spiral structure in the
inner Galaxy is to plot the integrated \hi\ emission as the lines of
sight sweep the Galactic equator in longitude.  This was done in the
\citet{burton71} paper, namely in Figure~12 there, for the
\hi\ emission observed along the Northern Galactic equator.
Figure~\ref{fig:bint} here shows the integrated emission scanned along
the Northern and the Southern quadrants of the disk.  This is done for
the total line of sight, but also by only integrating over gas in the
subcentral region.  In these plots, the valleys and hills that would
be expected as the lines of sight sweep in longitude crossing interarm
and arm regions of contrasting accumulated \hi\ material, low and
high, respectively, of do not reveal such structure.  The data are
``noisy'' but this is expected given feedback from the stellar winds
of massive stars and supernovae that will be encountered over the
lengths of path considered.  This again is a robust conclusion, not
dependent on distance measurements.

\citet{hou15} plot, in the bottom panel of their Figure~2, the
integrated intensity of \hi\ over the velocity range of $\pm
15$\kms\ from the ``terminal velocity'' which they define as the
maximum velocity expected for a flat rotation curve.  The authors
associate the bumps in this plot with spirals arms in the Milky Way.
But using a flat rotation curve means, faced with the rather messy
kinematic situation, that emission at more extreme velocities will be
smeared along the locus of subcentral points, guaranteeing an
elongated spiral-arm arch, and that regions with only weak emission
populating the lesser velocities corresponding to a flat rotation
curve will result in elongated interarm arch.  Their error made by
using a flat rotation curve---we do not need a rotation
curve---disqualifies the lower panel of their Figure 2.  This is the
analysis technique that lead, mistakenly, to the appearance of
majestic spiral structure in the Kootwijk map of 1957.  Our "terminal
velocity" is the extreme observed velocity, not a velocity which
corresponds to a flat rotation curve drawn through the ups and downs
of the kinematic edge of the longitude-velocity diagram.

We comment that the conclusion requires that the \hi\ ensemble be
optically thin; if it were opaque, the lower- and higher-density
regions would be obscured by rather flat-topped spectra.  But the
ensemble is not opaque, as proven by the ridge seen in
velocity-longitude diagrams near $V_{\rm LSR} = 0$\kms: the well-known
distance ambiguity is such that two regions are sampled at positive
velocities, with one region sampled at negative velocities; this for
the North, but analogously with velocities reversed for the South.  If
the optical depth of the disk layer were high, the spectra would be
rather flat-topped and the ridge near $V_{\rm LSR} = 0$\kms\ would not be
apparent.

Although the trends shown in the left and right portions of
Figure~\ref{fig:bint} are largely similar, there are differences of
details.  This is of no consequence to our conclusions: many galaxies,
including our own, are to some degree lopsided.  \citet{Sancisi08}
have reviewed this rather common situation.

By definition Grand-design spirals have strong arm-interarm contrast,
but what about a moderate contrast that can be observed in many
external spirals galaxies?  \citet{bittner17} investigated the
contrast between arm and interarm for Grand-design, Multi-armed, and
Flocculent galaxies.  They suggest that the Multi-armed type are
an intermediate case between Grand-design and Flocculent.
\citet{bittner17} find that Flocculent galaxies have significantly
lower arm-interarm contrast.

We therefore suggest that the Milky Way belongs to the category of
Flocculent spirals.

\section{Summary}\label{summary}

We have revisited the remark made by Shane \& Bieger-Smith in 1966,
that the \hi\ spectra populating the Galactic equator do not show
shallow shoulders at the velocities contributed by material along the
locus of subcentral distances from the Galactic center.  We have done
this using data from the HI4PI survey, which is certainly superior to
data from the Dwingeloo 25-m telescope in sensitivity and resolution
in angle and in velocity used in the earlier work.  We then redo the
plot made by Burton in 1971 which showed the shape of the
high-velocity wing of the HI spectra measured along the locus, but
using the modern data available for the Northern as well as the
Southern quadrants of the disk.  We also determine the integrated
\hi\ brightness temperature pertaining to gas near the subcentral
region.  The results confirm that there is no evidence for interarm
regions between spiral arms; without evidence for interarm regions
there is as a consequence no compelling evidence for spiral arms
dominating the inner-Galaxy morphology.  This result applies for the
Southern as well as the Northern portions of the Galactic disk
interior to the Solar orbit.  The result is robust and quantitative,
not depending on the awkward matter of measuring distances.

We suggest that the inner Galactic disk is a collection of loosely
clumped features, in an apparent state of disorder, which have been
studied as ``bones'', ``spurs'', ``rifts'', or ``patches,'' but which
have not been shown to arranged in large, connected, features, or thus
to partake in a clear spiral pattern.  It is not practical to try to
list all of the papers written in the decades since the 1957 work that
have tried to envision---to fit or to force---a wide range of material
in a template of a Grand-design spiral.  We suggest that it would be
productive to explore further the possibility that the Milky Way is a
member of the class of Flocculent spiral galaxies.

\begin{acknowledgments}

We are grateful to an anonymous referee for constructive comments.
The National Radio Astronomy Observatory is a facility of the National
Science Foundation operated under cooperative agreement by Associated
Universities, Inc. This research has made use of NASA's Astrophysics
Data System Bibliographic Services.
  
\end{acknowledgments}

\vspace{5mm}

\software{Astropy (Astropy Collaboration et al. 2013), Matplolib
  \citep{hunter07}, NumPY \& SciPy \citep{vanderwalt11}.}

\bibliography{ms}

\end{document}